\documentclass{article}

\usepackage{arxiv}

\usepackage[utf8]{inputenc} % allow utf-8 input
\usepackage[T1]{fontenc}    % use 8-bit T1 fonts
\usepackage{hyperref}       % hyperlinks
\usepackage{url}            % simple URL typesetting
\usepackage{booktabs}       % professional-quality tables
\usepackage{amsfonts}       % blackboard math symbols
\usepackage{nicefrac}       % compact symbols for 1/2, etc.
\usepackage{microtype} 
\usepackage{graphicx}
\usepackage[utf8]{inputenc} %unicode support
\usepackage[flushleft]{threeparttable}
\usepackage{graphicx}
\usepackage{amsmath}
\usepackage{graphics}
\usepackage{caption}
\usepackage{array}
\usepackage{multirow}
\usepackage{environ}
\usepackage{verbatim}
\usepackage{xcolor}
\usepackage{floatrow}
\usepackage{textcomp}
\usepackage{epstopdf}
\usepackage{dblfloatfix}

\NewEnviron{resizealign}{\sbox0{
		$\begin{matrix}\displaystyle\BODY\end{matrix}$}%
	\sbox1{$(\theequation)$}%
	\sbox2{\parbox{\dimexpr \wd0 + 2\wd1}%
		{\begin{align}\BODY\end{align}}}% for testing
	\noindent\resizebox{0.6\textwidth}{!}{\usebox2}%
}

\DeclareMathOperator*{\argmin}{arg\,min}

\title{Regional Registration of Whole Slide Image Stacks Containing Highly Deformed Artefacts}

\author{
Mahsa Paknezhad\\
\textbf{Sheng Yang Michael Loh}\\
\textbf{Hwee Kuan Lee}\\
Imaging Informatics Division\\
Bioinformatics Institute\\
A*STAR, 30 Biopolis, Matrix\\
Singapore, 138671\\
\texttt{mahsap@bii.a-star.edu.sg}
\And
Yukti Choudhury\\
\textbf{Min-Han Tan}\\
Lucence Diagnostics\\
217 Henderson Road\\
Singapore, 159555\\
Institute of Bioengineering and Nanotechnology\\
31 Biopolis, Nanos\\
Singapore, 138669\\
\And
Valerie Koh Cui Koh\\
\textbf{Puay Hoon Tan}\\
\textbf{John Yuen Shyi Peng}\\
\textbf{Yong Zhen Loy}\\
Singapore General Hospital\\
Outram Road\\
Singapore, 169608\\
\And
Timothy Tay Kwang Yong\\
\textbf{Hui Shan Tan}\\
\textbf{Ravindran Kanesvaran}\\
\textbf{Yongcheng Benjamin Tan}\\
\textbf{Min-Han Tan}\\
National Cancer Centre Singapore\\
11 Hospital Drive\\
Singapore, 169610\\
\And
\textbf{Weimiao Yu}\\
Institute of Molecular and Cell Biology\\
61 Biopolis Drive\\
Singapore, 138673\\
\And
Hwee Kuan Lee\\
National University of Singapore\\
Singapore, 119077\\
Singapore Eye Research Institute\\
20 College Road\\
Singapore, 169856\\
Image and Pervasive Access Lab\\
1 Fusionopolis Way\\
Singapore, 138632\\
}

\begin{document}
\maketitle
\begin{abstract}
\textbf{Motivation} %if any
High resolution 2D whole slide imaging provides rich information about the tissue structure. This information can be a lot richer if these 2D images can be stacked into a 3D tissue volume. A 3D analysis, however, requires accurate reconstruction of the tissue volume from the 2D image stack. This task is not trivial due to the distortions that each individual tissue slice experiences while cutting and mounting the tissue on the glass slide. Performing registration for the whole tissue slices may be adversely affected by the deformed tissue regions. Consequently, regional registration is found to be more effective. In this paper, we propose an accurate and robust regional registration algorithm for whole slide images which incrementally focuses registration on the area around the region of interest.

\textbf{Results} %if any
Using mean similarity index as the metric, the proposed algorithm (mean $\pm$ std: $0.84 \pm 0.11$) followed by a fine registration algorithm ($0.86 \pm 0.08$) outperformed the state-of-the-art linear whole tissue registration algorithm ($0.74 \pm 0.19$) and the regional version of this algorithm ($0.81 \pm 0.15$).  The proposed algorithm also outperforms the state-of-the-art nonlinear registration algorithm (original: $0.82 \pm 0.12$, regional: $0.77 \pm 0.22$) for whole slide images and a recently proposed patch-based registration algorithm (patch size 256: $0.79 \pm 0.16$ , patch size 512: $0.77 \pm 0.16$) for medical images.

\textbf{Availability}
The C++ implementation code is available online at the github repository: https://github.com/MahsaPaknezhad/WSIRegistration

\end{abstract}

% keywords can be removed
%\keywords{First keyword \and Second keyword \and More}

\section{Introduction}
\label{sec:introduction}
%What are the goals of using whole slide images?
Whole slide imaging has been widely used for education, digital archiving and teleconsultation since its development in 1999~\cite{ho2006use}. Despite the initial technical and logistical challenges such as the lack of universally accepted file formats, maintaining focus in an uneven sample, optimizing resolution, brightness and contrast, adoption of whole slide images (WSIs) for research and diagnosis has been gradually increasing~\cite{higgins2015app}. Clinical studies using WSIs and light microscopy have shown comparable accuracy for primary diagnostics in breast pathology~\cite{al2012digital}, gastrointestinal tract pathology~\cite{al2012whole}, and urinary system pathology~\cite{al2014whole}.\par
\begin{figure*}[!t]
    \centering
\includegraphics[clip=true, trim=0cm 15cm 5cm 2cm, width=\textwidth]{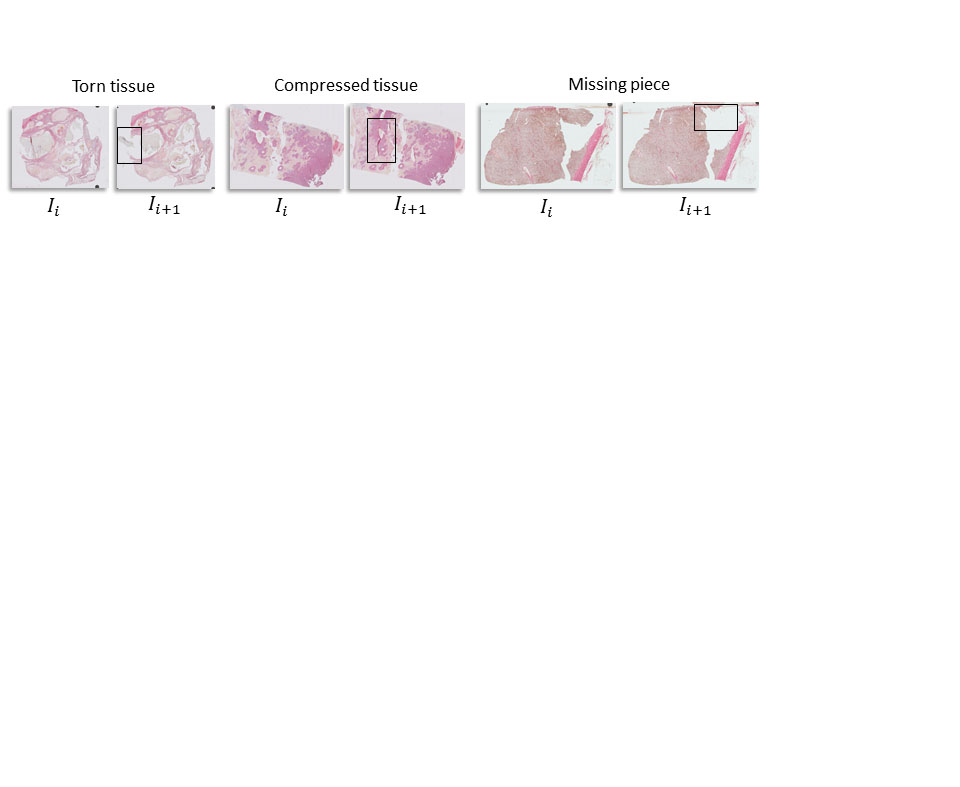}
\caption{A few examples of tissue deformation in two consecutive whole slide images ($I_i$, $I_{i+1}$). Such deformations make registration of whole slide images challenging.}
\label{fig:distortions}
\end{figure*}

Another advantage of WSIs is the ability to view and compare the same specimen with different stains and at different magnifications simultaneously. Diagnosis, prognosis and treatment planning is conducted by means of certain biomarkers that identify any normal or abnormal condition in the tissue. These biomarkers usually point to the structure of cells, tissues, genes and protein expressions and are captured using standard staining techniques. Gram staining and Immunohistochemistry (IHC) are typical examples of such techniques. Whole slide imaging has also allowed for developing automatic analysis and diagnostic tools for more accurate identification and study of diseases. High-resolution 3D reconstruction of the original tissue volume from the 2D slices of WSIs enables researchers to study certain features which cannot be revealed in the 2D images, such as the vascular structure, length and branching of vessels or colocalization of biomarkers. During the tissue cutting and mounting process, each individual thin section may experience certain deformations. As a result, an accurate 3D reconstruction of the tissue volume is not feasible without first registering the scanned tissue in subsequent virtual slides.\par

The image acquisition process for a small volume of fixed tissue involves cutting the volume into very thin sections and mounting them on microscope slides for staining. Scanning is preformed for each slide individually after staining. The morphological changes which may occur to the tissue during slide preparation such as tissue compression or stretching, missing or torn tissue, stain variations and artifacts, rotation and translation of the tissue~\cite{rastogi2013artefacts} are some of the reasons why registration of whole slide images is challenging. Fig.~\ref{fig:distortions} shows a few examples of tumor tissue samples from patients with clear cell renal cell carcinoma. Tissue compression, missing tissue and tearing are found very often in the scanned images. Currently, there exists no algorithm which can recover the highly deformed tissue regions shown with black boxes in Fig.~\ref{fig:distortions}. This is also not the aim of the proposed registration algorithm in this paper. However, these deformed regions may occupy a large part of the image, and therefore can adversely affect the registration results for other parts of the tissue in the image. A naive solution to this problem could be to crop the region of interest in the 2D image stack and perform registration for the cropped regions. However, these regions may not have enough image information based on which an accurate registration could be achieved. We propose a multi-scale registration algorithm for whole slide images where registration is initiated on a larger area around the region of interest and increasingly concentrates registration on a smaller area around the region of interest. The proposed algorithm provides accurate and robust regional registration results despite the existence of high tissue deformations in the images. It should be noted that no information regarding the location of the highly deformed regions is provided to the proposed image registration technique. \par   
\begin{figure*}[!t]
\centering
\includegraphics[clip=true, trim=0cm 1.5cm 0cm 0cm, width=0.85\textwidth]{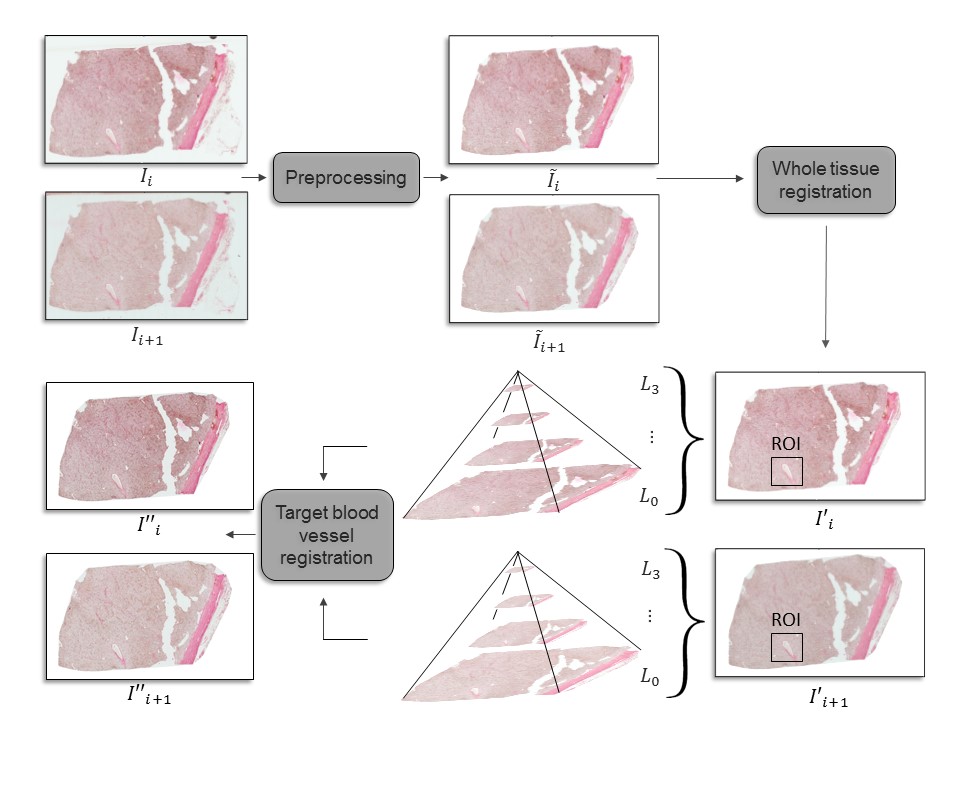}
\caption{The diagram shows an overview of the proposed algorithm for regional registration in whole slide images. The 'preprocessing' step removes the extra stains and artifacts around the tissue of interest. The 'whole tissue registration step' approximately aligns the whole tissue in consecutive whole slide images. Finally, the blood vessel of interest is registered in consecutive slides using a multi-scale approach in the 'target blood vessel registration' step. }
\label{fig:overview}
\end{figure*}
Due to the large size of whole slide images in their full resolution, well-established registration methods cannot be deployed to register these images with good accuracy without a high performance computing system. Application of these methods on lower resolutions may also result in significant registration errors in the full resolution. Currently proposed whole slide tissue registration algorithms take different approaches in order to keep the computation time reasonable as well as to find the global optimum solution. The majority of proposed algorithms for whole slide images can be classified into three approaches or a combination of them. The first class of methods are multi-scale approaches~\cite{wang2013improved,moles2014registration,lotz2014zooming}. In multi-scale registration methods, a coarse registration is carried out in the initial stage. The registration output is later refined by performing further registrations. In the work by Wang and Chen \cite{wang2013improved}, for instance, the images are sparsely represented in the coarse level by extracting SIFT features~\cite{lowe1999object}. The extracted features are used to find the optimal transformation between the two images. In the fine level, an area based b-spline method is deployed to improve the registration; while in the work by Moles Lopez et al. \cite{moles2014registration}, a four-level pyramidal registration with linear transformations is utilized. To increase the registration speed, similarity is only measured on randomly sampled pixels in this approach.\par
The second class of methods are patch-based approaches~\cite{roberts2012toward,song2014unsupervised,lotz2016patch}. In patch-based methods, the image is divided into regularly spaced patches and registration is carried out for each patch individually. The method proposed by Balakrishnan et al. \cite{balakrishnan2019voxelmorph,dalca2019unsupervised} is a good example of this class of algorithms. In this algorithm, the images are registered in an affine manner first. Later, a convolutional neural network is trained on randomly selected pairs of moving and fixed image patches from the training data set in an unsupervised manner to generate a deformation field that registers the moving image patch to the fixed image patch. Using a loss function that takes into account image similarity and deformation field smoothness, the trained network is able to register the whole image pairs in the test data set.  Many patch-based algorithms are also multi-scale. The work by Roberts et al. \cite{roberts2012toward}, is a good example, where consecutive slides are aligned nonrigidly first. Patch-based registration is then performed for increasing resolutions of the images. Finally, a global b-spline nonrigid transformation is estimated from the set of rigid patch transforms. This task is performed for increasing resolutions of the images with each output transformation initializing the transformation for the next resolution. The work by Lotz et al.~\cite{lotz2016patch} is another example of  multi-scale patch-based methods where nonrigid registration is carried out on the patches that are allowed to overlap. Vessel segmentation based algorithms~\cite{schwier2013registration,liang2015liver} are the third class of algorithms which take advantage of the vessel structure in the images to improve the registration output. In the work by Schwier at al.~\cite{schwier2013registration}, for example, vessels are extracted from each image and nonrigid registration is performed using the vessel masks. The work by Liang et al.~\cite{liang2015liver} combines all the three approaches and proposes a multi-scale registration algorithm which rigidly aligns the patches, fuses the rigid transformations by a cubic b-spline deformation and performs vessel segmentation and association later on to reduce the registration error.\par
%Why do we need registration?
Some analysis that need 3D volumes and cannot be carried out on 2D slices are as follows: 1) Recent studies investigate the role of Pericyte cells, the cells detected around the walls of capillaries and micorvessels, in cancer progression. Both Endothelial and Pericyte cells are found to undergo morphological and architectural abnormalities~\cite{morikawa2002abnormalities}. Hamzah et al. \cite{hamzah2008vascular} suggest that Pericyte cells aid in tumor evasion of immune rejection. It is therefore of great interest to study the role of Pericyte cells in tumor angiogenesis for which a high-resolution 3D reconstruction of blood vessels is necessary in order to colocalize the Pericyte cells in the tumor volume. 2) Reconstruction of the tissue volume can help differentiate imaged glands that are cut tangentially during the tissue slicing process from small poorly formed glands, and therefore helps in more accurate cancer diagnosis. 3) Detection of small lesions and immune cell infiltration will be easier in a high resolution 3D reconstructed tissue volume. 4) Blood vessel structure can also be used to extract important features about blood vessels such as branching of vessels, length of vessels, etc.\par
Considering the high frequency of tissue deformations in the acquired virtual slices and the large size of these images, previously proposed algorithms are not able to provide an accurate tissue registration output in a reasonable amount of time. Also, patch-based (\cite{roberts2012toward,song2014unsupervised,lotz2016patch}) and vessel-segmentation-based methods (\cite{schwier2013registration,liang2015liver}), discussed earlier, may be misguided by highly deformed regions in the tissue. We propose a multi-scale approach for registration of the whole slide images for the region of interest while reducing the effect of such distortions on the registration results and therefore increasing the accuracy of the registration outcome. In this paper, we have focused on registration of the tissue around blood vessels in order to be able to quantify the registration accuracy using manual lumen segmentations. The next section, provides details on different steps of the proposed algorithm. 

\section{Implementation}
In order to register a target blood vessel in the whole slide images, three steps are carried out as follows: 1) Preprocessing, to remove extra stains and artifacts around the tissue of interest, 2) Whole tissue registration, to approximately align the whole tissue in consecutive whole slide images, 3) Target blood vessel registration, to register the blood vessel of interest. Finally, fine registration is carried out to improve the registration for the blood vessel of interest. Fig.~\ref{fig:overview} gives an overview of the steps of the proposed method. A flowchart for each step is provided in the supplementary materials.
\subsection{Preprocessing}
Extra stains and artifacts around the tissue of interest can affect the registration outcome. To remove these effects, each image ($I_i$) is converted to the gray scale and smoothed using a Gaussian filter with a standard deviation of 10 pixels. The smoothed image is then thresholded using the image's mean pixel intensity as the threshold value. As a proper segmentation of the tissue cannot be achieved merely by thresholding, the undetected tissue regions such as the boundary areas are recovered by applying a closing and later an opening morphological operation on the thresholded image using a circular kernel of radius 20 pixels. Contours in the opened image are then detected. The contour(s) which is (are) closer to the center of the image and surround(s) the largest area in the image is (are) identified. Extra tissue and stains outside the convex hull of the selected contour(s) are removed, resulting in a cleaned tissue image ($\tilde{I}_i$). In the next step, registration of the whole tissue is carried out for consecutive whole slide images.
\subsection{Whole tissue registration}
\label{sec:wholetissueregistration}
In this stage, any relative rotation or displacement in the location of the tissue is corrected. The cleaned image, $\tilde{I}_i$, is segmented using the multi-resolution Monte Carlo method of Sashida et al.~\cite{sashida2016application} which performs piece-wise constant segmentation of the Mumford and Shah~\cite{mumford1989optimal} model to yield $\tilde{I}_i^{MS}$. Mumford-Shah segmentation of the image removes the noise, texture and small spatial intensity variations making the image clean and the upcoming registration robust against inter-slice intensity variations due to differences in stain densities. The reason for choosing Sashida's approach for segmentation of the Mumford-Shah model is mainly because of its outperformance over other approaches such as the work by Song and Chan~\cite{song2002fast} and Bae and Tai~\cite{bae2009graph} in multi-phase segmentation of images. Next, each consecutive pair of Mumford-Shah segmented images are registered independently. For each pair of images $\{\tilde{I}_i^{MS},\tilde{I}_{i+1}^{MS}\}$, a combination of varying translation $(dx,dy)$ and rotation ($\theta$) transformations are applied to the second (moving) image to find the rotation and translation parameters which make $T_{\theta,dx,dy}[\tilde{I}_{i+1}^{MS}(x,y)]$ most similar to $\tilde{I}_i^{MS}(x,y)$ by optimizing the following function: 
\begin{equation}
\argmin_{\{\theta,dx,dy\}}\sum_{x,y}\left(T_{\theta,dx,dy}[\tilde{I}_{i+1}^{MS}(x,y)]-\tilde{I}_{i}^{MS}(x,y)\right)^2
\end{equation}
The $\{\theta, dx, dy\}$ triplet which gives the least sum of squared difference is chosen and its corresponding transformation matrix is applied to the moving image: ${I'}_{i+1} \leftarrow  T_{\theta,dx,dy}[\tilde{I}_{i+1}]$. These two steps roughly align the images in consecutive image slides. Other algorithms which can be used for rough registration of the whole tissue in subsequent image slices include landmark-based registration methods or b-spline registration algorithms. Landmark-based methods may not provide a good initial registration in case the selected landmark is located in a tissue section which experiences severe deformations in certain image slices. Moreover, optimization of the parameters for a b-spline transformation technique is computationally demanding and inaccurate for images with strong distortions. Therefore, a linear transformation technique was utilized in this step. In the next step, the blood vessel of interest is registered in consecutive image slices. 
\subsection{Target blood vessel registration}
Registration of the whole tissue provides a good initial alignment for further registration of the target blood vessel. To decrease the influence of tissue deformations on registration of the blood vessel of interest, a regional registration approach is deployed at this stage. Different steps of this approach are explained below:
\begin{figure*}[!t]
\centering
\includegraphics[clip=true, trim=0cm 8.5cm 0.2cm 0cm, width=0.9\textwidth]{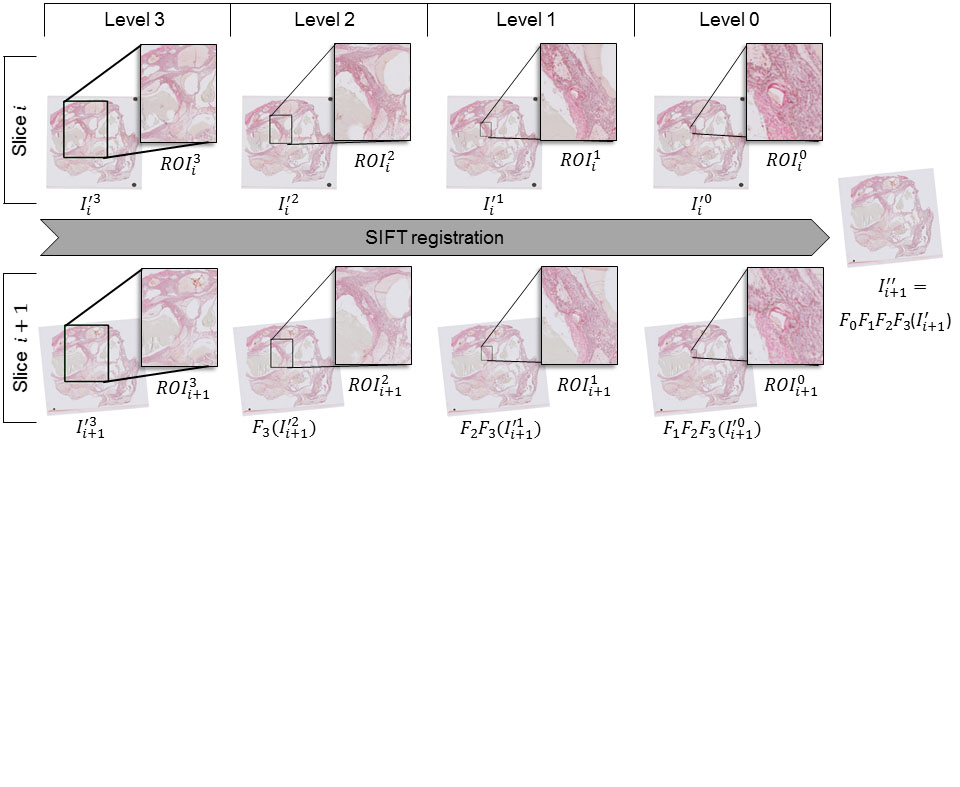}
\caption{Figure shows the region of interest (ROI) in multiple resolutions (levels 0 to 3 with level 0 referring to the highest resolution of the slices) of two consecutive whole slide images. The user defines the $ROI^0_i$ for the target image in its highest resolution ($I^0_i$). The ROI in lower resolutions ($ROI^{1,2,3}_i$) are defined automatically. $F_r$ refers to the best rigid transformation matrix found to align slice $i+1$ to slice $i$ in resolution level $r$ of the slices.}
\label{fig:siftregistration}
\end{figure*}
\subsubsection{Multi-resolution ROI extraction} Let us define the registered image sequence from the previous step as ${I'}_{seq} = \{{I'}_1, {I'}_2, \ldots\}$. A small area (a Box) around the blood vessel of interest is defined by the user for the image at its full resolution ($r=0$) as $ROI^{r=0} \equiv ROI^{0}$  (see $ROI_i^{0}$ in Fig~\ref{fig:siftregistration}). In order to register $ROI^{0}$ in ${I'}_i$ and ${I'}_{i+1}$ image slices, registration is first performed for lower resolutions of the two images. ${I'}_i^{r=k}$ ($\equiv {I'}_i^{k}$) is denoted as the output image from downsampling ${I'}_i$ by a factor of $2^k$. Similarly, a downsampled image is generated for image ${I'}_{i+1}$ and is denoted as ${I'}_{i+1}^{k}$. Registration is performed between ${I'}_{i}^{k}$ and ${I'}_{i+1}^{k}$ by considering $ROI^{k}$ only. The same procedure is taken for $r = \{k-1, k-2, \ldots, 0\}$ in the presented order giving a series of rigid transformations ($F_{r}$) as follows:
\begin{small}
\begin{align}
   SIFT_{{ROI}^k}\left({I'}_{i}^{k}, {I'}_{i+1}^{k}\right)& \rightarrow  & F_{k} \nonumber\\[5pt]
   SIFT_{{ROI}^{k-1}}\left({I'}_{i}^{k-1}, F_{k}\left({I'}_{i+1}^{k-1}\right)\right) & \rightarrow &  F_{k-1}F_{k} \nonumber\\[5pt]
   \vdots \nonumber\\[5pt]
    SIFT_{{ROI}^{0}} \left({I'}_{i}^{0}, F_{1}\dots F_{k}\left({I'}_{i+1}^{0}\right)\right) & \rightarrow & F_{0} \ldots F_{k-1} F_{k}
    \label{eq:sift}
\end{align}
\end{small}
where $SIFT_{ROI} (I,J)$ performs registration for the specified $ROI$ in the image pair $(I,J)$ using SIFT features and will be explained in Section~\ref{sec:siftregistration}. The resulting transformations are applied to image ${I'}^0_{i+1}$. The ROI for different image resolutions is defined as follows: If $(x_{ROI},y_{ROI})$ is the center coordinates of $ROI^{0}$ in ${I'}_i^{0}$, the corresponding center coordinates in the downsampled image ${I'}_i^{k}$ are calculated as $(x_{ROI}/2^k,y_{ROI}/2^k)$. The ROI ($ROI^{k}$) for ${I'}_i^{k}$ is then extracted with the same width and height as for $ROI^{0}$ unless the ROI boundaries exceed the image boundaries. In that case, the width and height are adjusted such that the ROI stays in the image boundaries. Fig.~\ref{fig:siftregistration} shows the extracted ROI from different image resolutions for the manually identified ROI ($ROI^0_i$) in the full resolution image (${I'}^0_i$). A significant difference between the proposed method and the existing methods lies in this step, where registration is initiated in a larger field of view around the ROI in the lowest resolution and incrementally focuses more on the registration of the ROI in higher resolutions. Moreover, the multi-resolution nature of the algorithm makes it computationally efficient. Next section describes the $SIFT$ function in Eq.~\ref{eq:sift} or how registration of the $ROIs$ is performed in each resolution. 
\subsubsection{Registration using SIFT features}
\label{sec:siftregistration}
Having extracted the region of interest $ROI^k$ in the lowest resolution images ${I'}^{k}_i$ and ${I'}^{k}_{i+1}$, distinctive key points are detected in both ROIs using SIFT feature extraction method~\cite{lowe1999object}. Spatial coordinates of the detected key points are added to the descriptors of the key points and are taken into account for key point matching in the two ROIs. If more than 8 good matches are found for the two ROIs, the top 8 matches are selected, otherwise all the matches are selected. Since registration is performed locally, a rigid registration is found sufficient. A rigid transformation can be calculated with a minimum of 3 key points per image. Therefore, $\binom{8}{3}=56$ different combinations of 3 matches and consequently, 56 different transformation matrices can be obtained using the 8 selected matches. Our experiments show that these combinations provide a sufficiently diverse set of transformations from which an accurate registration can often be achieved.  All the transformations are applied to the previously registered image ${I'}_{i+1}^{k}$ in Section~\ref{sec:wholetissueregistration}  giving a series of 56 warped images
$
    \{F_1\left({I'}_{i+1}^{k}\right), F_2\left({I'}_{i+1}^{k}\right), \ldots, F_{56}\left({I'}_{i+1}^{k}\right)\}
$
where $F_m$ is the transformation matrix found by using a unique combination of 3 matches. From the 56 transformation proposals, the transformation matrix which gives the smallest sum of squared difference in pixel intensity ($D$) for $ROI_k$ in the warped image $F_m\left({I'}_{i+1}^{k}\right)$  and the reference image ${I'}_{i}^{k}$ is chosen using the following function:
\begin{equation}
F_k  =  \displaystyle\min_{F_m} D_{ROI_k}\left({I'}_{i}^{k}, F_m({I'}_{i+1}^{k})\right), \forall m \in [1, 56]\\
\end{equation}

The best transformation matrix found for resolution $k$ is then scaled up and applied to the moving image in the higher resolution ${I'}_{i+1}^{k-1}$. This process is repeated for resolutions $r = \{k-1, k-2, \ldots, 0\}$ in the presented order and the registration for the image pair (${I'}_{i}$, ${I'}_{i+1}$) is finalized by defining the transformation matrix for $r=0$ as $ F^*  =  F_0 F_1 \ldots F_k$ and applying it to the original moving image: ${I''}_{i+1}  \leftarrow  F^*({I'}_{i+1})$. Similarly, registration is performed for the registered image ${I''}_{i+1}$ and the next image in the stack (${I''}_{i+1}$, ${I'}_{i+2}$) aligning the blood vessel of interest in the entire image stack: ${I''}_{seq} = \{{I''}_1, {I''}_2, \ldots, {I''}_{i+1}, {I''}_{i+2}, \ldots\}$.  A diagram of this process is shown in Fig.~\ref{fig:siftregistration} (Bottom). In the next section, we evaluate the proposed method on a data set of whole slide image stacks acquired from clear cell renal cell carcinoma patients.

\section{Evaluation and Results}
\label{results}
Our dataset consists of whole immunohistochemical slide images of three patients with clear cell renal cell carcinoma. All three specimens were fixed in formalin and embedded in paraffin blocks. A sequence of 100 slices were cut with image resolution of 0.5 $\mu$m/pixel for patient \#1, and 0.25 $\mu$m/pixel for patient \#2, and 150 slices were cut with image resolution of 0.25 $\mu$m/pixel for patient \#3 from the tissue blocks with 4  $\mu$m thickness. All tissue sections were double-stained to reveal the endothelial cell marker CD31 (Dako M0823, Clone JC70A, 1:50 dilution, Epitope Retrieval 2, pH 9.0) and the pericyte marker $\alpha$-SMA Dako M0851, Clone 1A4, 1:1000 dilution, Epitope Retrieval 2, pH 9.0). Immunohistochemical staining was performed using the Leica Bond Max autostainer (Leica Biosystems Melbourne) according to the manufacturer’s instructions. The double stains were visualized using the Bond Polymer Refine Detection and the Bond Refine Red Detection systems (Leica Biosystems), with CD31 staining brown and SMA staining red. Each slide was scanned at high resolution with IntelliSite Pathology Ultra Fast Scanner (Philips Digital Pathology Solutions, The Netherlands) and viewed with IMS Viewer (Philips, The Netherlands). All registration experiments were performed on level 4 magnification of the whole slide images as the full resolution image in this paper. Although our data set consists of merely three acquired image stacks, a combination of 20 blood vessels with different shape and orientation and from different regions in the images were selected for registration to adequately test robustness of the proposed algorithm. Four resolution levels were considered for the proposed registration algorithm. The lumen for each blood vessel was segmented manually. The output transformation matrices for the registered blood vessels were applied to the corresponding lumen segmentations to measure the registration accuracy. It is important to emphasize that the lumen segmentations were solely used to compare the performance of the proposed method with the competing methods. These segmentations were not used in any steps of the registration pipeline. \\

\begin{table}
\caption{Table shows the set of parameter values that were tested for the multi-resolution registration algorithm proposed by Moles Lopez et al.~\cite{moles2014registration}. The parameter values which resulted in the best mean registration accuracy for all the 20 blood vessels are shown in bold.}
    \centering
       \begin{tabular}{>{\centering}p{3cm}>{\centering}p{3cm}>{\centering}p{0.1cm}}
       \hline\\
        Parameter & All resolutions & \\[5pt]
       \hline\\
        $T_\mu$ & \textbf{Affine}, Rigid&\\[5pt]
        \hline\\
        S & \textbf{MI}, NCC&\\[5pt]
        \hline\\ 
        $N_s$ &  \textbf{8000} &  \\[5pt]
        \hline\\ 
        $N_i$ &  \textbf{2000} &\\[5pt]
        \hline\\
        MSL & 1, 5, \textbf{10}, 20 & \\[5pt]
        \hline\\ 
        Input & \textbf{Lum}, Hem & \\[5pt]
        \hline
		\end{tabular}
		\label{tab:parameters}
\end{table}
A 2D view of the blood vessels that were selected for registration are shown in the first column in Fig.~\ref{fig:vessels1} for 10 blood vessels. 
In the supplementary materials, we have provided the registration results for a few slides of these blood vessels, the best SIFT feature matches for the ROIs in the previous and current slides, and the selected 3 SIFT feature matches that resulted in the best alignment of the 2 slides are also provided. We applied the proposed algorithm on all 20 selected blood vessels. The 3D reconstruction of the selected blood vessels before registration, and the resulting 3D reconstructed blood vessels after performing registration using the proposed algorithm are shown in the second and third column, respectively, in Fig.~\ref{fig:vessels1} and Fig.~\ref{fig:vessels1}. The original image scale in the z dimension is compressed for presentation purposes. As can be seen, the vessels are well-aligned using the proposed algorithm for most of the slices except for a few slices pointed out by the red boxes.\par
In order to have a quantitative measurement of the accuracy of the proposed registration algorithm we used similarity index as the metric. Similarity index measures the similarity between the non-zero pixels in the two lumen masks after registration using the following formula: 
\begin{equation*} \nonumber
S = 2\frac{|A \bigcap B|}{|A| + |B|}
\end{equation*}
Where A and B are the set of non-zero pixels in the first and the second lumen mask, respectively. The operator $|.|$ defines the size of the set and the operator $\bigcap$ represents the intersection of the two sets. This metric was chosen for quantitative analysis mainly because similarity index stays meaningful for any two consecutive images in which blood vessel branching occurs. The similarity index ($mean\pm std$) for lumen segmentations after registration using the proposed algorithm for the 20 blood vessels and 5 consecutive slices was measured and is reported as (Proposed Algorithm) in Table~\ref{tab:quantitativeComparison}. \par
We compared our algorithm with the state-of-the-art multi-scale registration method by Moles Lopez et al.~\cite{moles2014registration} mentioned in Introduction Section where they introduce a four-level registration algorithm. Similar to our approach, their algorithm performs registration on the lowest-resolution image first. The resulting deformation field is applied to the next higher-resolution image before performing registration on this level. In order to make their algorithm faster and more robust, they measure the similarity metric only on randomly sampled pixels in the image. They take the stochastic gradient descent (SGD) method as the optimizer and consider only linear transformations ,$T_\mu$, (affine and rigid) for registration. This algorithm is evaluated using different metrics and input parameters such as Mattes mutual information (MI) and normalized cross correlation (NCC) as the similarity metric, multiple number of pixels to evaluate the similarity metric ($N_s$), different number of iterations, $N_i$, of the SGD optimization procedure, and two different channels extracted from whole slide images, namely, the hematoxylin channel (blue) and the luminance channel.\par Table~\ref{tab:parameters} shows the set of input parameters for this algorithm. The parameter values were taken from the set of parameters evaluated for high-resolution image experiments in the paper by Moles Lopez et al.~\cite{moles2014registration} (Table 1 in the paper) except for the values for maximum step length (MSL) where we found the range {1, 5, 10, 20} outputting more accurate registration results. We tested the algorithm with different combination of the parameter values defined in Table ~\ref{tab:parameters} for 5 consecutive slices and found the set of values that gave the best mean registration accuracy for all the 20 blood vessels to be $T_\mu = Affine$, $S = MI$, $MSL = 10$, and $Input = Lum$.\\
\begin{table}[!t]
\caption{Mean registration accuracy of the lumen segmentations after registration. The accuracy was measured for lumen segmentations of 20 blood vessels for 5 consecutive slices after registration using the method proposed by Moles Lopez et al.~\cite{moles2014registration} (Moles Lopez et al. \cite{moles2014registration} 1 Round), the regional version of this method (Moles Lopez et al. \cite{moles2014registration} 2 Rounds), the method proposed by Wang and Chen~\cite{wang2013improved} (Wang and Chen~\cite{wang2013improved} - 1 Round), the regional version of this method (Wang and Chen~\cite{wang2013improved} - 2 Rounds), and the patch-based method proposed by Balakrishnan et al.~\cite{balakrishnan2019voxelmorph} with patch sizes 256 $\times$ 256 pixels and 512 $\times$ 512 pixels. Their accuracy was compared with those of the proposed algorithm, and the proposed method followed by fine registration using Moles Lopez et al.~\cite{moles2014registration}.}
    \begin{threeparttable}
		\begin{tabular}{>{\centering}p{6cm}>{\centering}p{2.5cm}>{\centering}p{3.7cm}p{0cm}}
         \hline\\
         Method &  Similarity index & Exec. time & \\
         \hline\\[5pt]
         Moles Lopez et al. \cite{moles2014registration} - 1 Round &  $0.74 \pm 0.19$  & 5.8 min * &\\[5pt]
         \hline\\
         Moles Lopez et al. \cite{moles2014registration} - 2 Rounds &  $0.81 \pm 0.15$ & 6.4 min * &\\[5pt]
         \hline\\ 
         Wang and Chen \cite{wang2013improved} -  1 Round & $0.82 \pm 0.12$ & 2.8 min $\dagger$ &\\[5pt]
         \hline\\ 
         Wang and Chen \cite{wang2013improved} - 2 Rounds & $0.77 \pm 0.22$  & 3.0 min $\dagger$ &\\[5pt]
         \hline\\ 
        \multirow{2}{*}{Balakrishnan et al. \cite{balakrishnan2019voxelmorph} - Patch size 256} & \multirow{2}{*}{$0.79 \pm 0.16$} &  80.5 min (for training) *&\\
         & & 0.34 min (for testing) *&\\[5pt]
         \hline\\ 
         \multirow{2}{*}{Balakrishnan et al. \cite{balakrishnan2019voxelmorph} - Patch size 512} & \multirow{2}{*}{$0.77 \pm 0.16$}  &  316.9 min (for training) *&\\
         &  &       0.35 min (for testing) *&\\[5pt]
         \hline\\ 
         Proposed Algorithm &   $0.84 \pm 0.11$ & 3.4 min * &\\[5pt]
         \hline\\
         Proposed Algorithm followed   &  \multirow{2}{*}{$\mathbf{0.86 \pm 0.08}$} &  \multirow{2}{*}{5.6 min *} & \\
         by Moles Lopez et al.~\cite{moles2014registration}& & &\\[5pt]
         \hline 
    \end{tabular}
    \begin{tablenotes}
        \small
        \item
       * Ubuntu 19.04.4 LTS 64-bit, Intel Core i7-6700 CPU 3.40 GHz x 8, 31.4GB RAM $\dagger$ Windows 8.1 Pro 64-bit, Intel Core i7-4720HQ CPU 2.60GHz, 11.9GB RAM
    \end{tablenotes}
    \end{threeparttable}
    \label{tab:quantitativeComparison}
\end{table}
Using this set of values, we applied the method on the preprocessed images generated by our algorithm ($\tilde{I}_{seq}$) for all the 20 blood vessels. The mean registration accuracy for this method is reported as Moles Lopez et al. \cite{moles2014registration} 1 Round in Table.~\ref{tab:quantitativeComparison}. To have a fair comparison, the method proposed by Moles Lopez et al.~\cite{moles2014registration} was also provided with the same manual user input ($ROI_i^0$). The ROI defined by user ($ROI_i^0$) was extracted from the registered images and further registration was performed on the cropped regions. This approach is referred to as (Moles Lopez et al. \cite{moles2014registration} 2 Rounds) in Table~\ref{tab:quantitativeComparison}. After applying the proposed registration algorithm we extracted the ROI from the registered images and performed another round of registration using the method by Moles Lopez et al.~\cite{moles2014registration} to see if any improvements in the registration accuracy can be achieved. We refer to this method as proposed method followed by fine registration throughout the paper. Registration accuracy for the proposed algorithm followed by fine registration using the work of Moles Lopez et al.~\cite{moles2014registration} is also reported in Table~\ref{tab:quantitativeComparison}.\\ 
\begin{figure}[!t]
        \centering
		\includegraphics[clip=true, trim=5.5cm 1cm 1.5cm 0.5cm, width=0.8\textwidth]{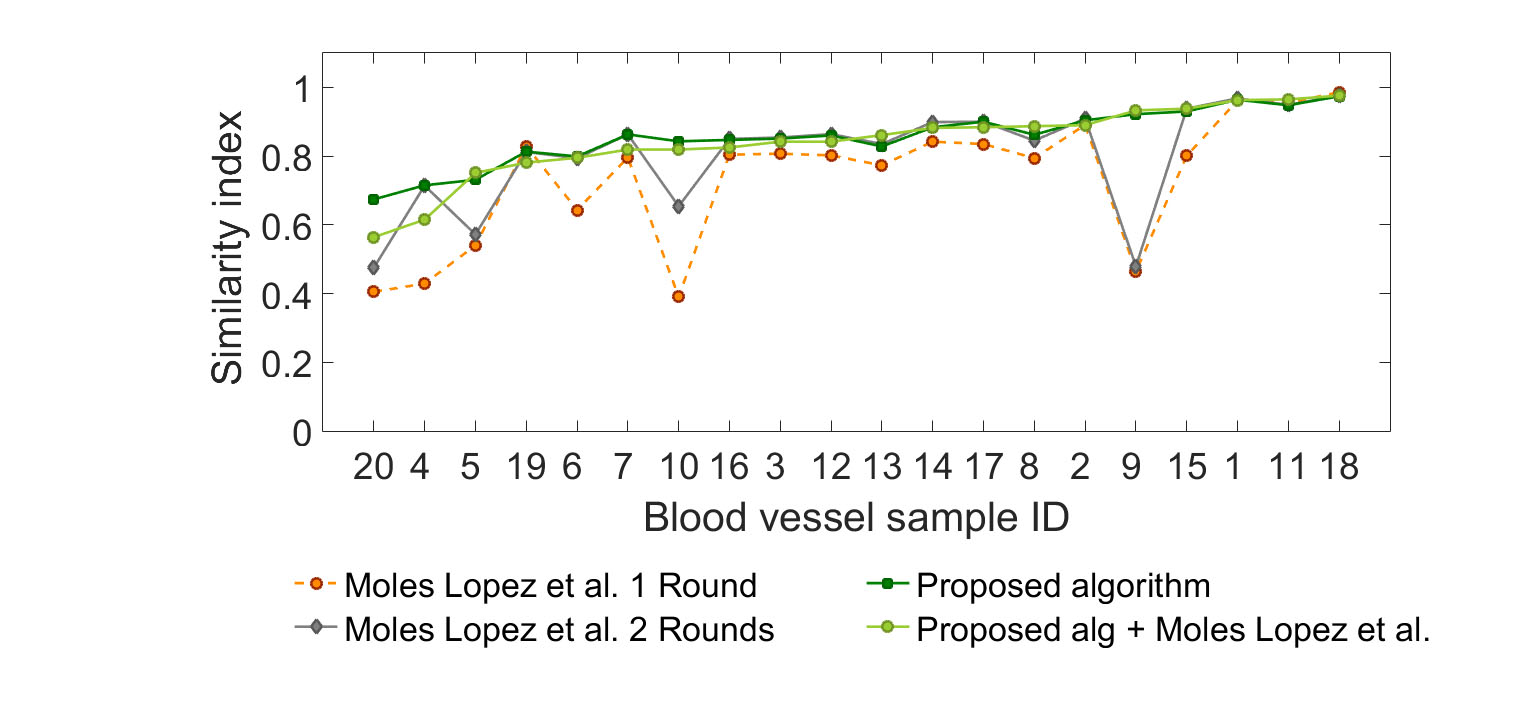}
		\caption{Figure compares registration accuracy using the proposed method, the proposed+fine registration (Proposed alg \&  Moles Lopez et al. \cite{moles2014registration}), the original (Moles Lopez et al. \cite{moles2014registration} 1 Round) and the regional version (Moles Lopez et al. \cite{moles2014registration} 2 Rounds) of the method by Moles Lopez et al.~\cite{moles2014registration} for 5 consecutive slices.}
		\label{fig:compare-ml-5}
\end{figure}
\begin{figure*}[!t]
\centering
\begin{tabular}{c}
      \includegraphics[clip=true, trim=5.5cm 1cm 1.5cm 1.1cm, width=0.8\textwidth]{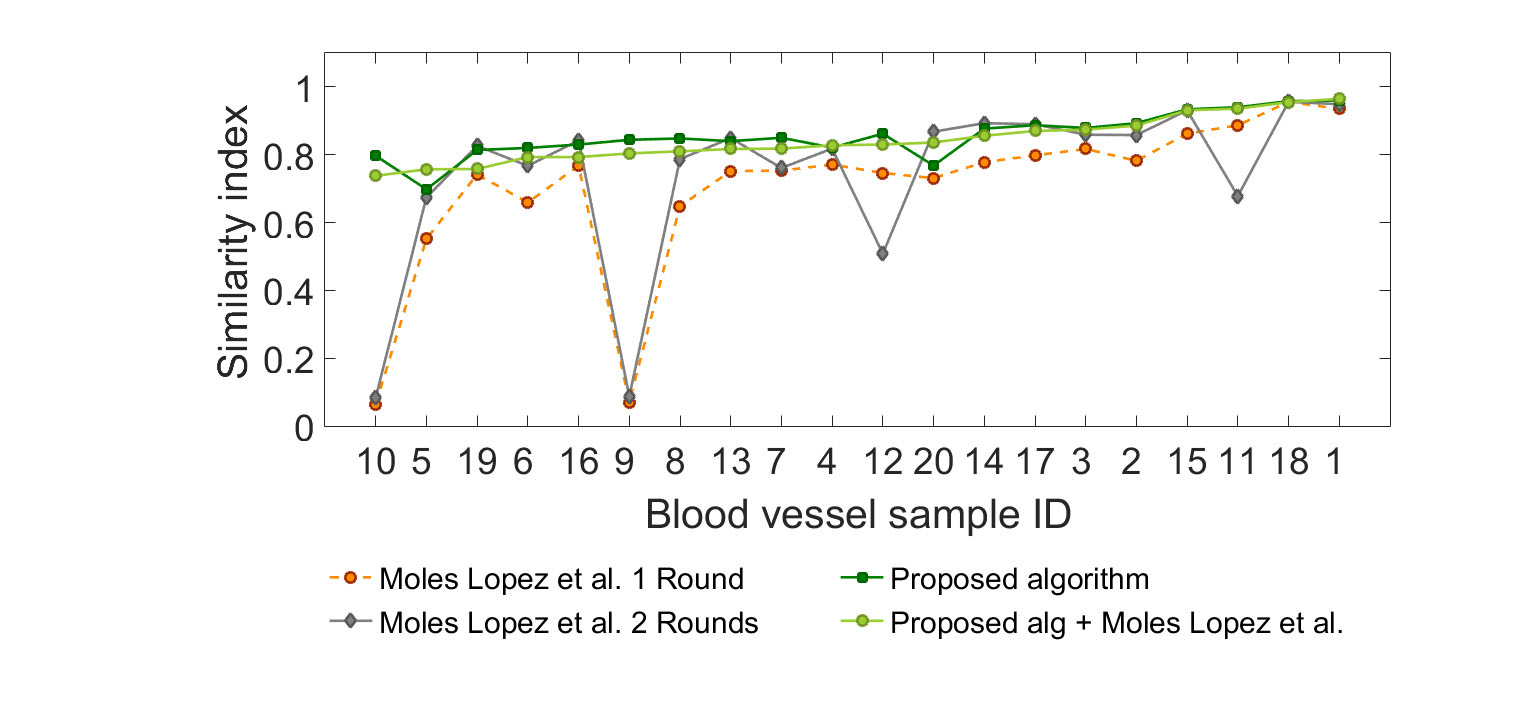}\\
      \includegraphics[clip=true, trim=5.5cm 1cm 1.2cm 1.1cm, width=0.8\textwidth]{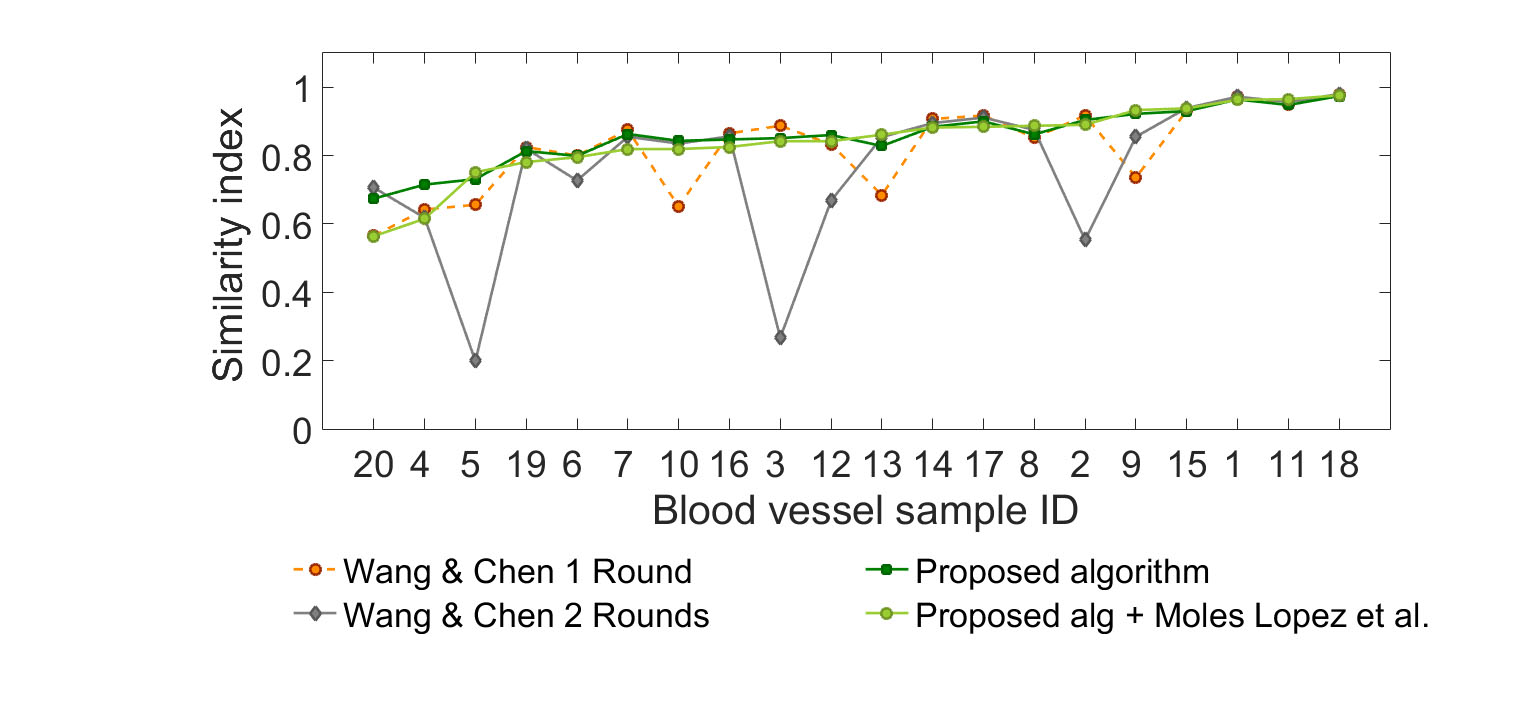}\\
\end{tabular}
      \caption{(Top) compares the similarity index measured for the proposed method and the method proposed by Moles Lopez et al.~\cite{moles2014registration} applied on all slices of the tissue volume (100-150 slices). (Bottom) compares our results with the original (Wang and Chen \cite{wang2013improved} 1 Round) and the regional version (Wang and Chen \cite{wang2013improved} 2 Rounds) of the method by Wang and Chen~\cite{wang2013improved} for 5 consecutive slices.}
      \label{fig:compare-ml-150-compare-wang-5}
\end{figure*}
Fig.~\ref{fig:compare-ml-5} compares registration accuracy of the proposed method and the proposed method followed by fine registration using the method by Moles Lopez et al. \cite{moles2014registration} with those of the whole tissue registration method (Moles Lopez et al. \cite{moles2014registration} 1 Round) and the regional version of this method (Moles Lopez et al. \cite{moles2014registration} 2 Rounds) for each blood vessel for 5 consecutive slices. As shown, both the proposed method and the proposed$+$fine registration provide more accurate and robust results. We also performed registration for all tissue slices (100-150) for the 20 blood vessels using the proposed algorithm and compared the results with the three other methods in Fig.~\ref{fig:compare-ml-150-compare-wang-5} (Top). \par
\begin{figure*}[t]
\begin{tabular}{c}
      \includegraphics[clip=true, trim=7cm 0cm 3cm 1.1cm, width=0.8\textwidth]{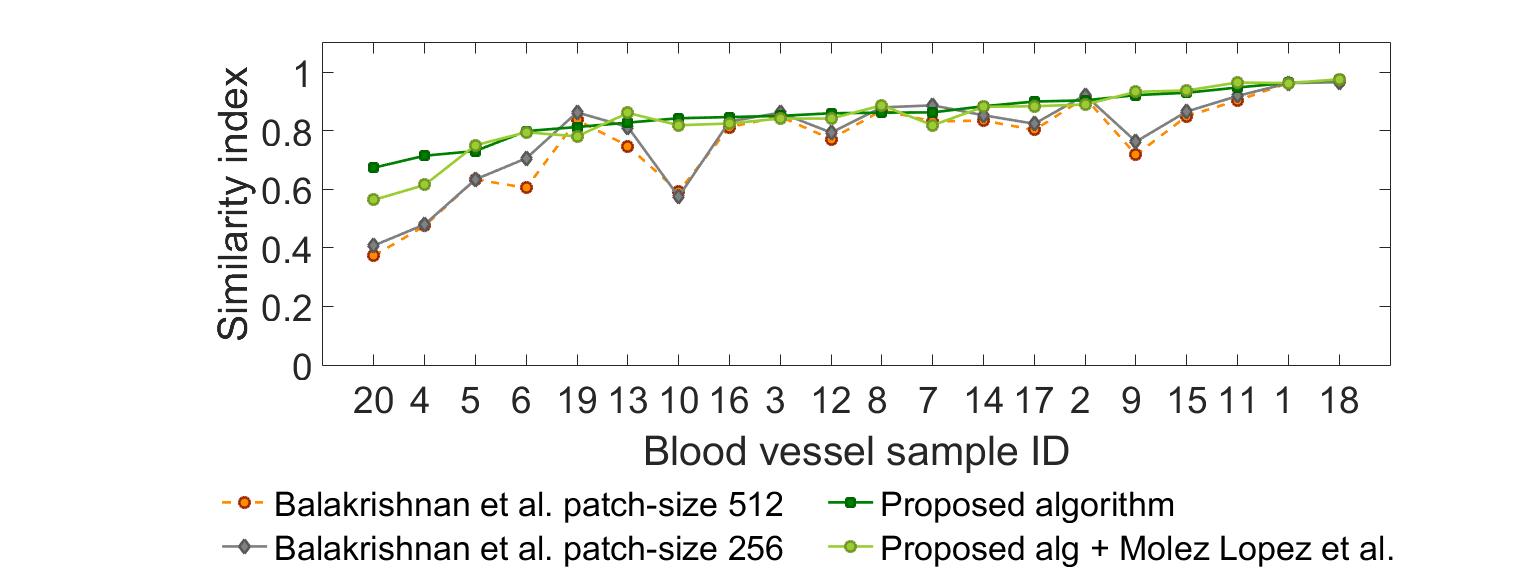}\\
      \includegraphics[clip=true, trim=7cm 1cm 3cm -0.5cm, width=0.8\textwidth]{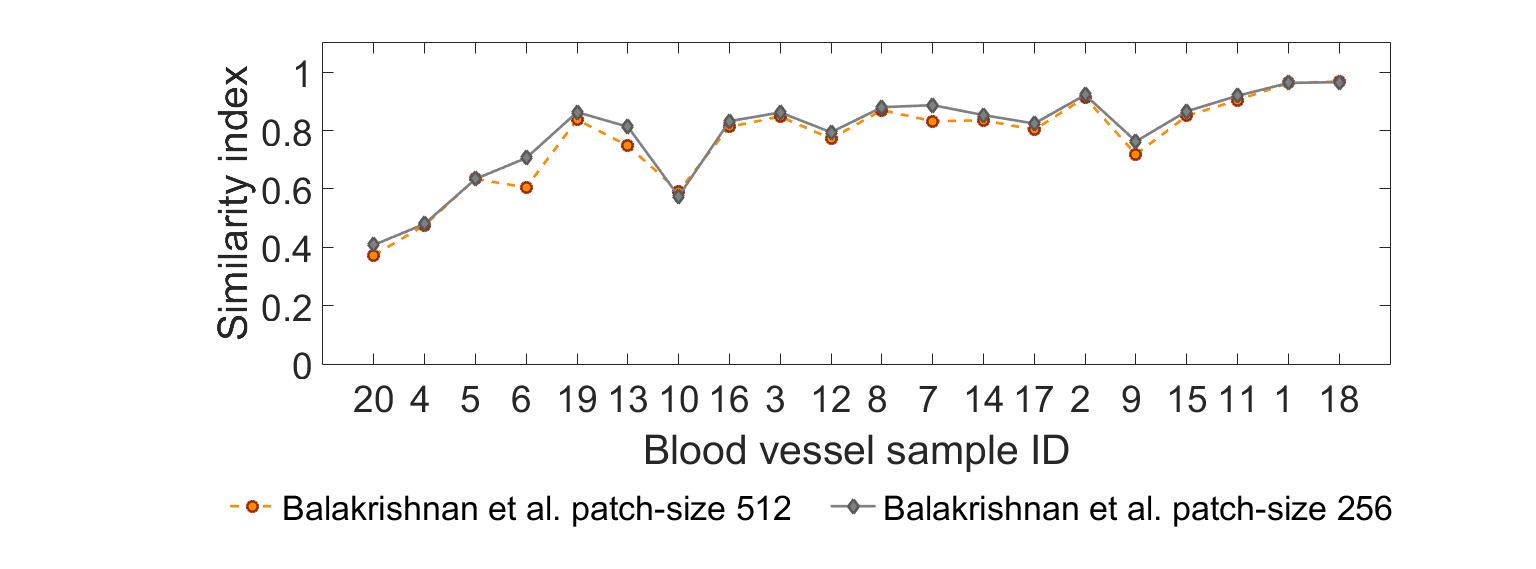}\\
\end{tabular}
      \caption{(Top) compares the similarity index measured for the proposed method and the method proposed+fine registration (Proposed alg \&  Moles Lopez et al. \cite{moles2014registration}), and the patch-based registration method by Balakrishnan et al.~\cite{balakrishnan2019voxelmorph} with patch sizes 256 and 512. (Bottom) shows the registration results for the patch-based registration method \cite{balakrishnan2019voxelmorph} with patch sizes 256$\times$256 and 512$\times$512.}
      \label{fig:compare-ml-dalca-5-compare-dalca-5}
\end{figure*}

 
The measured similarity index for the proposed algorithm and the proposed$+$fine registration algorithm was $0.84 \pm 0.07$, and $0.86 \pm 0.06$, respectively, both outperforming the work of Moles Lopez et al.~\cite{moles2014registration} ($0.70 \pm 0.24$) and the regional version of this algorithm ($0.74 \pm 0.25$) for all 100-150 slices. The fourth and fifth column in Fig.~\ref{fig:vessels1}  show 3D reconstruction of the registered lumen masks using the proposed algorithm followed by fine registration, and regional version (Moles Lopez et al. \cite{moles2014registration} 2 rounds) of the method by Moles Lopez et al.~\cite{moles2014registration} as the registration technique. \\
We also compared our algorithm to the patch-based registration algorithm proposed by Balakrishnan et al. \cite{balakrishnan2019voxelmorph}. We first registered all slices in an affine manner using the Fiji tool~\cite{schindelin2012fiji}. Later, we trained the proposed neural network model and applied the trained model on the images from the same patient to have a fair comparison with the other methods. In order to register 5 consecutive image slices for each patient, we first trained the proposed convolutional neural network model in an unsupervised manner on randomly selected patches cropped from these image slices. The trained model was then applied on the same data set to register consecutive pairs of whole slide images. We trained two models using patch-sizes of 256 $\times$ 256 and patch-sizes of 512 $\times$ 512 for 30 iterations with a batch-size of 8 and 500 pairs of patches sampled in each iteration. The regularization parameter value for the loss function was set to 1, a learning rate of $1e-3$ was defined and mean squared error was utilized as the image similarity metric. The results for both patch-sizes are compared with our proposed registration results in Fig.~\ref{fig:compare-ml-dalca-5-compare-dalca-5} (Top). The registration results for the patch-based registration algorithm by Balakrishnan et al.~\cite{balakrishnan2019voxelmorph} for different patch-sizes can be more easily seen in Fig.~\ref{fig:compare-ml-dalca-5-compare-dalca-5} (Bottom). As can be seen, the patch-based registration algorithm is not able to provide a good registration of the blood vessels of interest compared to our proposed method as the registration is affected by other regions on the whole slide image.  
In order to ensure that the proposed registration algorithm and the final registration results after applying fine registration are robust against different sizes of the region of interest (ROI), we considered different ROI sizes around the blood vessels of interest and performed registration using the proposed algorithm and the competing methods. The results are provided in the supplementary materials. The results suggest that performing regional registration after applying the proposed algorithm does not always provide reasonable results for the small-sized ROIs for the experiments with $MSL = 10$. However, regional registration is robust against medium- and large-sized ROIs with $MLS=10$ and different ROI sizes with $MSL = 1$.\\
We also compared the proposed algorithm to spline-based deformable registration method by Wang and Chen \cite{wang2013improved} mentioned in Introduction Section which, similar to our algorithm, performs multi-scale registration using SIFT features. In their method, however, the whole tissue is considered for registration. In order to improve the registration results, they perform an area-based bi-directional elastic b-spline registration as the final stage of their pipeline. In a similar way, we performed registration for the 20 defined blood vessels for 5 consecutive slices using the proposed method by Wang and Chen \cite{wang2013improved} (Wang and Chen \cite{wang2013improved} 1 Round) and the regional version of this method (Wang and Chen \cite{wang2013improved} 2 Rounds) and compared the results with those of our proposed method in Fig.~\ref{fig:compare-ml-150-compare-wang-5} (Bottom). Quantitative comparison is also provided in Table~\ref{tab:quantitativeComparison}. As can be seen, our algorithm outputs more robust results with better or similar accuracy supporting our assumption that a rigid registration is sufficient for local registration of the tissue.  
% TO BE ADDED Moreover, the multi-resolution nature of the algorithm makes it computationally efficient. 
%%TO BE ADDED** Performing registration for a large field of view (the low-resolution image) to a small field of view (high-resolution image) makes the registration process to incrementally focus more on the registration of the region of interest accurately.
The third column in Table~\ref{tab:quantitativeComparison} reports the measured execution time for registration of 5 consecutive slices by the corresponding algorithm. Note that the measured time for the proposed algorithm, the proposed algorithm followed by fine registration using Moles Lopez et al. \cite{moles2014registration}, the original and regional version of the proposed method by Moles Lopez et al. \cite{moles2014registration} were measured on a Ubuntu 19.04 LTS 64 bit system with Inter Core i7-6700 CPU 3.40 GHz x 8 with 31.4GB RAM, while the execution time for the original and regional version of the method by Wang and Chen \cite{wang2013improved} was measured on a Windows 8.1 Pro 64-bit system with Intel Core i7-4720HQ CPU 2.60GHz and 11.9GB RAM for technical reasons. The execution time reported in the Table for the patch-based method by Balakrishnan et al.~\cite{balakrishnan2019voxelmorph} refers to the average time spent to train the model using randomly extracted patches from the input image slides. It should also be mentioned that the measured time does not include the amount of time required by the physician for selecting the ROI for each blood vessel of interest. However, this task is trivial and usually takes a few seconds for the following reason. Selection of the ROI is performed in a lower resolution of the whole tissue slide such that the physician is able to see the blood vessels clearly and at the same time have a complete view of the whole tissue. The selected ROI by the physician in this resolution is multiplied by the ratio by which the image was downsampled for presentation to find the ROI for the full resolution of the image. It is important to note that the ROI that will be defined by our algorithm for this resolution will not be equivalent to the ROI defined by the physician mainly because our algorithm defines an ROI with the same size as the ROI size in the full resolution for all other resolutions unless the ROI boundaries exceed the image boundaries in the downsampled image in the current registration level. \\

\section{Discussion}
\label{sec:discussion}
An important novelty of the proposed registration algorithm compared to the previously proposed multi-scale registration algorithms is in the fact that registration is initiated for a large area (not necessarily the whole image) around the blood vessel of interest and increasingly concentrates on a smaller area (not the higher resolution of the whole image) around the target blood vessel as the resolution increases, giving more priority to a smaller area around the blood vessel of interest as registration approaches lower levels in the registration pipeline. This approach makes the algorithm desirable for registering a region of interest in high resolution images in a reasonable amount of time. Another contribution is in the way SIFT key points are utilized to  suggest a set of transformation matrices and find the best transformation matrix from this set.\par 
Moreover, our experiments approve the fact that simple rigid transformation models can result in better registration results even in the presence of nonrigid deformations. In other words, the transformation field for a nonrigidly deformed tissue can be approximated with rigid deformations which have taken place on small regions on the tissue.\par
The proposed method is staining-invariant and can be applied on multi-stained, double-stained, or Luminance images. The ability to perform the registration on full resolution of images increases accuracy of the results. Our experiments showed that the proposed method outperforms two state-of-the-art rigid and non-rigid registration algorithms.\par 
Although the proposed method was not evaluated on different image modalities in this paper, deploying this method for different modalities is straight forward as the only parameter that needs to be tuned is the number of layers in each octave for the SIFT feature detection algorithm which effects the number of detected key points on the tissue. For our experiments, we used the same number of layers per octave (10) for all the patient scans. \par
Despite the importance of registering whole slide images before performing tissue analysis, the work on whole slide image registration is quite limited. In this work, we tried our best to cite the most recent works on whole slide image registration and to compare our algorithm with the state-of-the-art methods. However, these methods are not very recent and do not provide the desired accuracy and robustness. This issue emphasizes the importance of developing more accurate and robust registration methods for whole slide images.  \par
An important issue that is left to address is how to merge the acquired deformation fields for each tissue region to get a single global deformation field for the whole tissue while keeping the deformations smooth at borders of the small tissue regions. Existing patch-based registration methods such as the works by Roberts et al.~\cite{roberts2012toward} and Liang et al.~\cite{liang2015liver} provide solutions to this problem. However, they do not take into account the existence of highly deformed regions for which the registration results may be considerably poor. It is important to mention that while our proposed method performs better in existence of highly deformed regions around the region of interest, similar to the previously published methods, it does not output accurate registration results for highly deformed regions as the regions of interest. Therefore, merging deformation fields of the regions of interests in existence of highly deformed regions needs to be addressed. To make the algorithm more deployable, an automatic or semi-automatic method for detection of the ROIs would be desirable. Finally, measures should be taken to speed up the proposed method for near real-time applications. 

\begin{table*}[!t]
\centering
\caption{First column in figure shows a 2D view of the blood vessels that were chosen for registration. The second column shows 3D construction of this blood vessel before registration. The 3D reconstruction of the blood vessels after registration using the proposed algorithm, the proposed algorithm followed by Moles Lopez et al.\cite{moles2014registration}, and Moles Lopez et al. \cite{moles2014registration} 2 Rounds are shown in the third, fourth, and fifth column, respectively.}
\includegraphics[width=\textwidth]{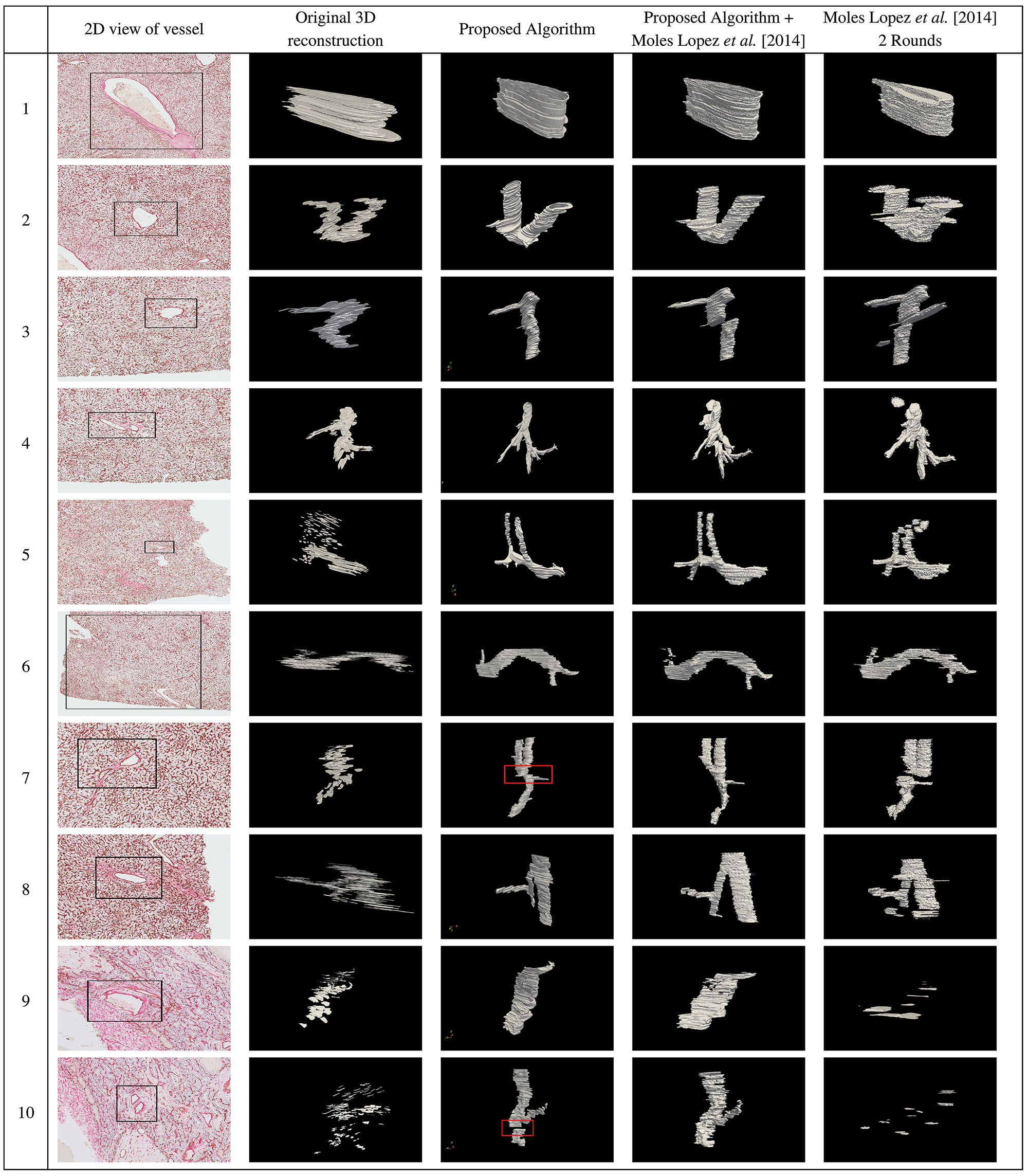}
\label{fig:vessels1}
\end{table*}

\begin{table*}[!t]
\centering
\caption{First column in figure shows a 2D view of the blood vessels that were chosen for registration. The second column shows 3D construction of this blood vessel before registration. The 3D reconstruction of the blood vessels after registration using the proposed algorithm, the proposed algorithm followed by Moles Lopez et al.\cite{moles2014registration}, and Moles Lopez et al.\cite{moles2014registration} 2 Rounds are shown in the third, fourth, and fifth column, respectively.}
\includegraphics[width=\textwidth]{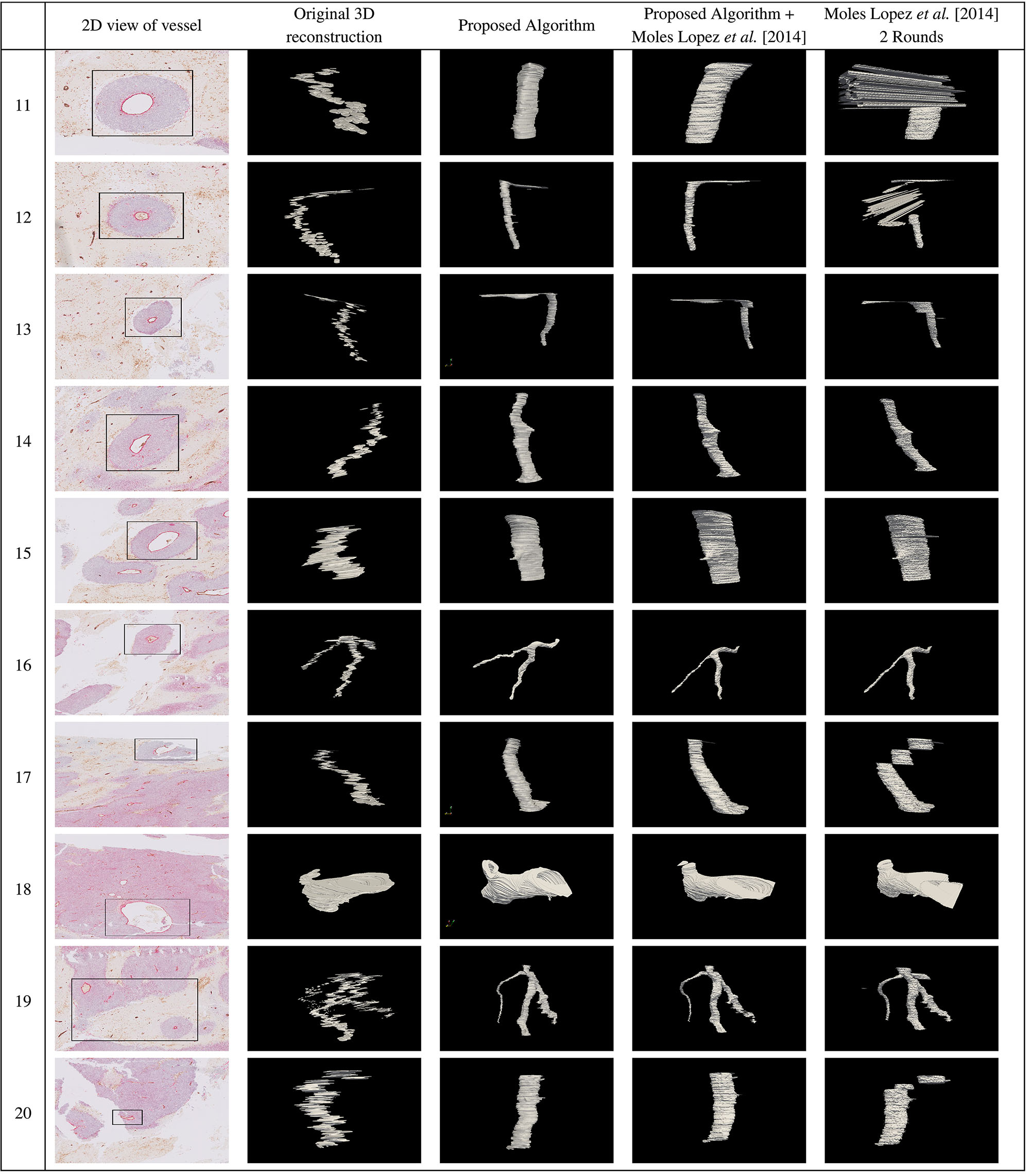}
\label{fig:vessels2}
\end{table*}

\section{Conclusion}
In order to study the tumor tissue in terms of vasculature and cell population, 3D reconstruction of the sliced and imaged tumor volume is necessary. We proposed a regional and multi-scale registration algorithm which provides more robust and accurate results compared to two linear and nonlinear whole slide image registration algorithms and one patch-based medical image registration algorithm. The better registration accuracy of our proposed method can be attributed to the fact that registration is incrementally focused on a smaller field of view around the tissue region of interest in higher resolutions while giving a global solution. Moreover, the proposed method needs only minor parameter tuning. Future work includes analysis of the reconstructed tissue volumes by the proposed algorithm to study drug influence on angiogenesis and cell populations in tumors.

%%%%%%%%%%%%%%%%%%%%%%%%%%%%%%%%%%%%%%%%%%%%%%
%%                                          %%
%% Backmatter begins here                   %%
%%                                          %%
%%%%%%%%%%%%%%%%%%%%%%%%%%%%%%%%%%%%%%%%%%%%%%

\section*{Acknowledgements}
  The authors would like to thank Dr. Adrian V. Dalca for his constant support that allowed us to apply the proposed algorithm by Balakrishnan et al. \cite{balakrishnan2019voxelmorph} on our data set. 
%%%%%%%%%%%%%%%%%%%%%%%%%%%%%%%%%%%%%%%%%%%%%%%%%%%%%%%%%%%%%
%%                  The Bibliography                       %%
%%                                                         %%
%%  Bmc_mathpys.bst  will be used to                       %%
%%  create a .BBL file for submission.                     %%
%%  After submission of the .TEX file,                     %%
%%  you will be prompted to submit your .BBL file.         %%
%%                                                         %%
%%                                                         %%
%%  Note that the displayed Bibliography will not          %%
%%  necessarily be rendered by Latex exactly as specified  %%
%%  in the online Instructions for Authors.                %%
%%                                                         %%
%%%%%%%%%%%%%%%%%%%%%%%%%%%%%%%%%%%%%%%%%%%%%%%%%%%%%%%%%%%%%
\bibliographystyle{unsrt} 
\bibliography{Main}

\end{document}

\maketitle

\begin{figure*}[!h]
\centering
\includegraphics[clip=true, trim=0.5cm 0cm 1.2cm 1.4cm, width=0.8\textwidth]{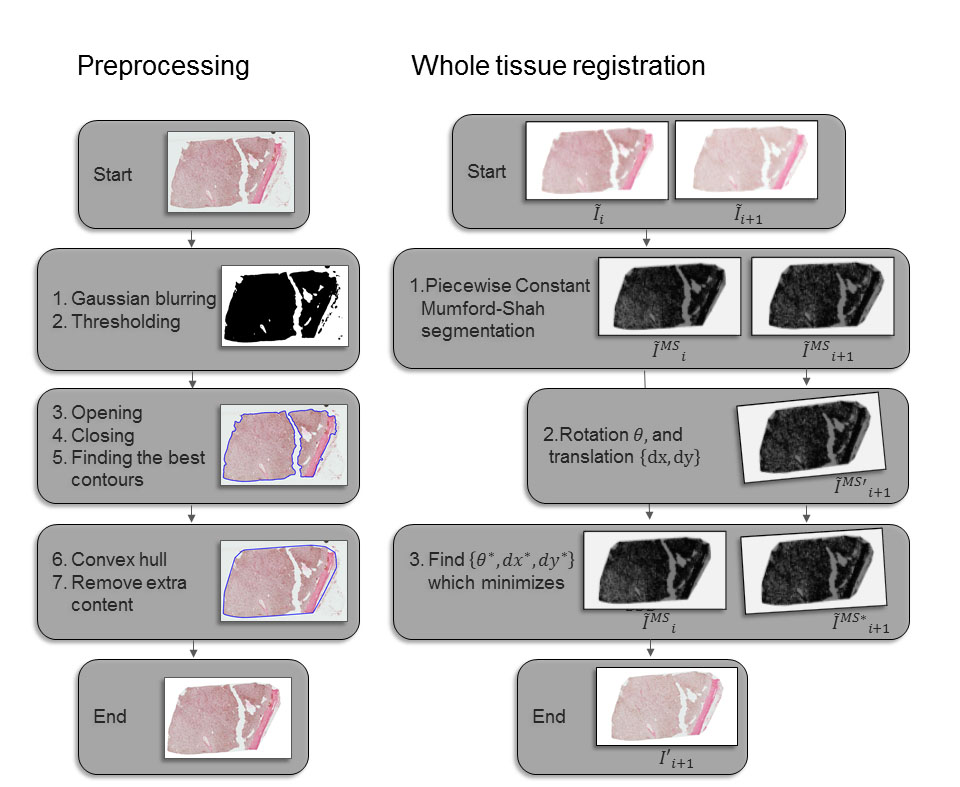}
\caption{Figure shows in more details the steps that are carried out in the 'Preprocessing' and 'Whole tissue registration' stages of the proposed algorithm.}
\label{fig:prep-and-ms-registration}
\end{figure*}

\begin{figure*}[!h]
\centering
\includegraphics[clip=true, trim=0cm 0cm 1.5cm 0cm, width=\textwidth]{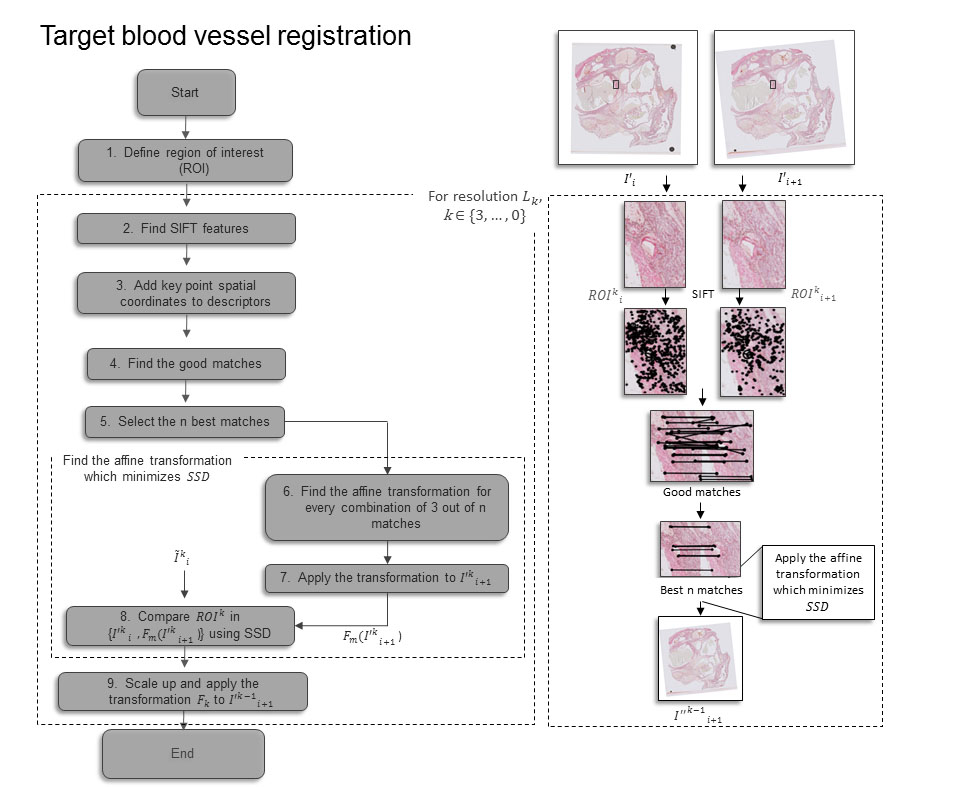}
\caption{ Figure shows the steps that are taken in the 'SIFT registration' stage of the algorithm to register the ROI for subsequent whole slide images.}
\label{fig:siftregistration}
\end{figure*}

Fig. (\ref{fig:prep-and-ms-registration})  shows the flowchart for the 'preprocessing' and 'whole slide tissue registration' step of the proposed algorithm and Fig. (\ref{fig:siftregistration}) shows the flowchart for the SIFT registration step of the algorithm. 

\clearpage 

\clearpage

\begin{figure*}[!h]
\includegraphics[clip=true, trim=0cm 0cm 0cm 0cm, width=\textwidth]{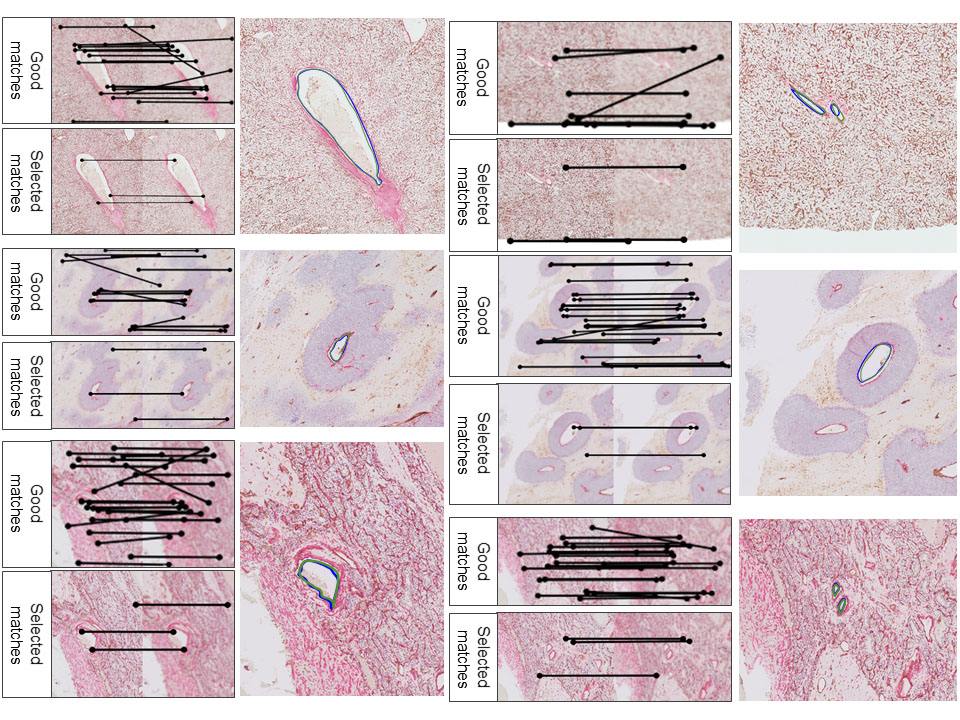}
\caption{Example of the SIFT registration results. For each example, the left top figure shows the matched SIFT features for the full resolution ROI in the previous and current slide ($ROI^0_{i,i+1}$). The bottom left figure shows the 3 SIFT matches that led to the best alignment between the two slides. The right figure shows segmentation for the previous slide (blue) mapped on the current slide after performing registration. Also shown is the segmentation for the current slide (green). One can see that the two segmentations are reasonably aligned.}
\label{fig:siftexamples}
\end{figure*}

Fig.~\ref{fig:siftexamples} shows the SIFT registration results for a few slides of the registered blood vessels. For each example, the top left image shows the best SIFT feature matches for the ROIs in the previous and current slides. Also shown is the selected 3 SIFT matches that resulted in the best alignment of the 2 slides. The image on the right shows the segmentation for the previous slide (blue) mapped on the current slide after performing registration. The segmentation for the current slide is also shown in green. One can see that the two slides are aligned fairly well.\par

\begin{figure*}[t]
     \begin{tabular}{cc}
        \includegraphics[clip=true, trim=3cm 4cm 0cm -2cm, width=0.5\textwidth]{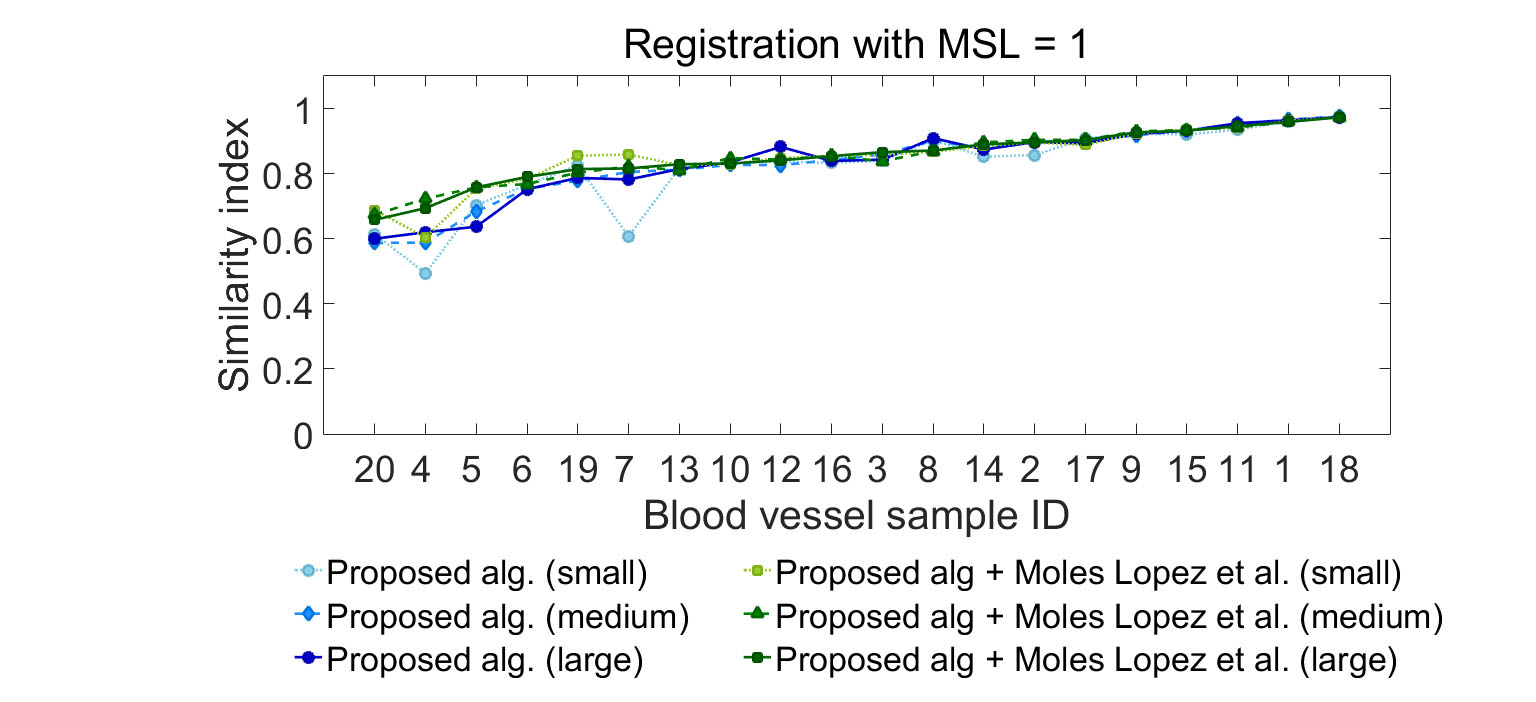}&
        \includegraphics[clip=true, trim=3.2cm 4.4cm 0cm -2cm, width=0.5\textwidth]{MSL1ProposedMLVsML2Rounds.jpg}\\
        \includegraphics[clip=true, trim=3cm -1.5cm 0cm -2cm, width=0.5\textwidth]{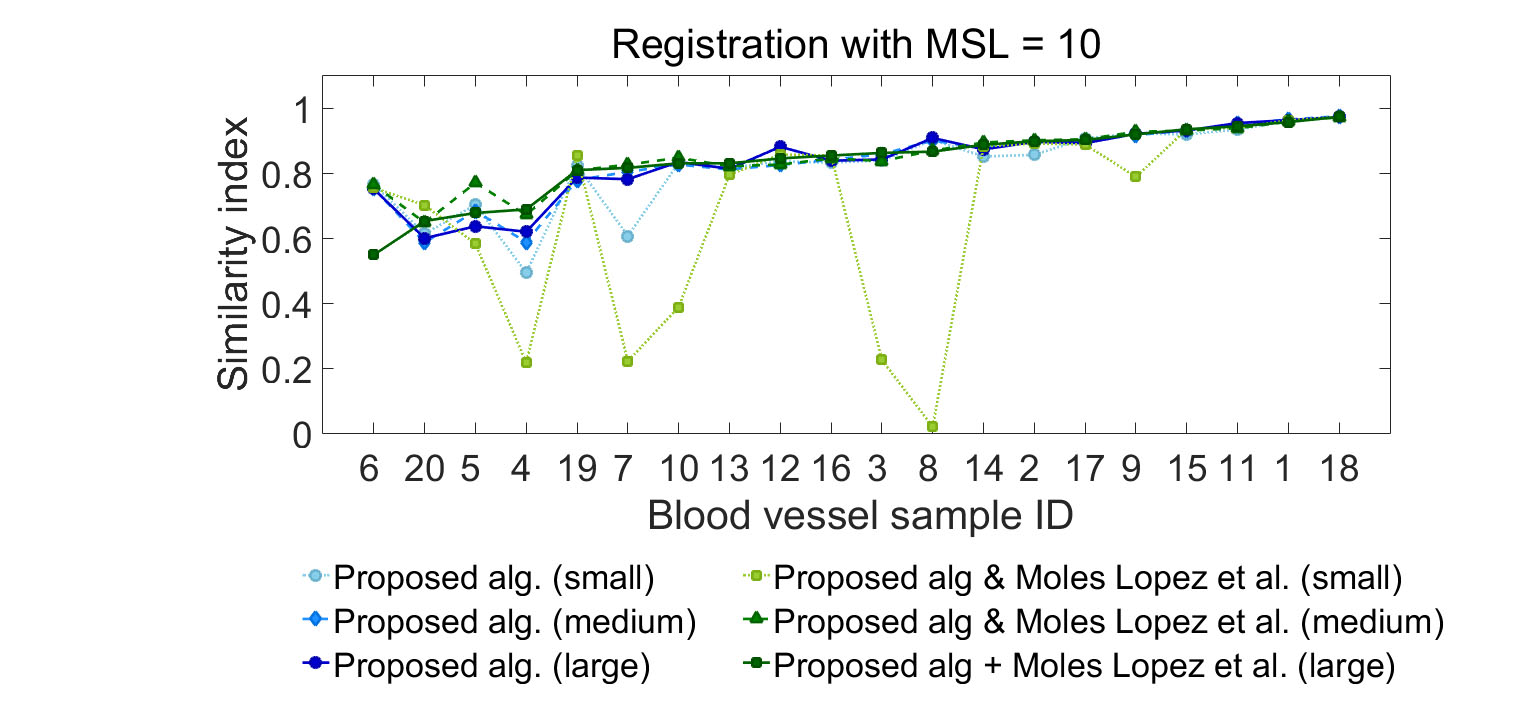}&
        \includegraphics[clip=true, trim=3cm -1.5cm 0cm -2cm, width=0.5\textwidth]{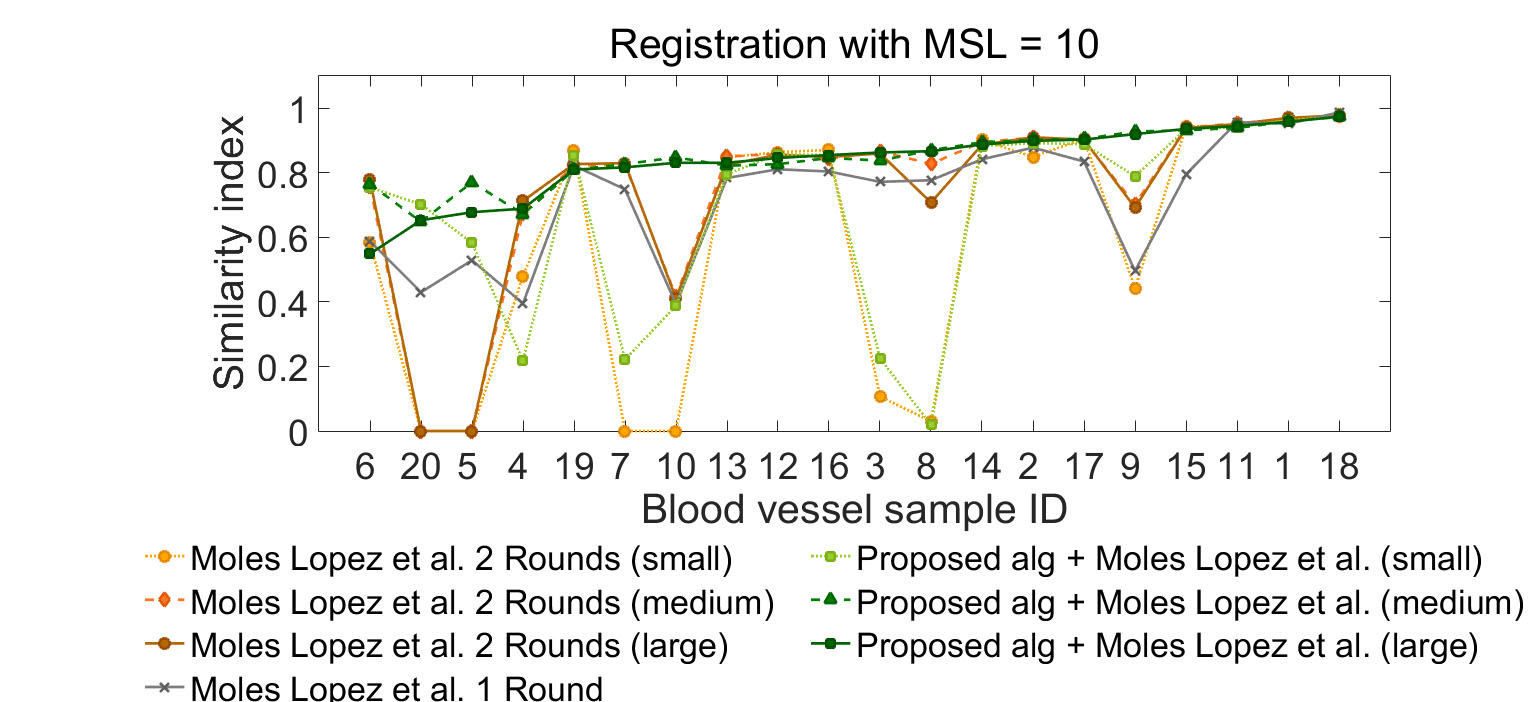}\\
      \end{tabular}
      \caption{(Left) shows the registration accuracy using the proposed registration algorithm using different ROI sizes. (small) and (large) refer to the smallest and largest ROIs considered for the blood vessels, respectively (with (large) ROI being 150 pixels bigger than the (small) ROI in both dimensions). Proposed alg. + \cite{moles2014registration} shows the registration accuracy for the proposed registration algorithm followed by fine registration using different ROI sizes. (Right) compares the registration accuracy with those of the competing algorithms (\cite{moles2014registration} 1 Round and 2 Rounds). (Top row) shows the results using $MSL = 1$ for fine registration.  (Bottom row) shows these results for fine registration using $MSL = 10$.}
      \label{fig:differentrois}
\end{figure*}

\clearpage
In order to ensure that the proposed registration algorithm and the final registration results after applying fine registration are robust against different sizes of the region of interest (ROI), we considered three different ROI sizes around the blood vessels of interest and performed registration using the proposed algorithm and the competing methods. After applying the proposed algorithm, fine registration was carried out using the work of~\cite{moles2014registration} with the best identified parameter values: $T_\mu = Affine$, $S = MI$, $MSL = 10$, and $Input = Lum$ and another set of values with smaller MSL: $T_\mu = Affine$, $S = MI$, $MSL = 1$, and $Input = Lum$. As shown in Fig.~\ref{fig:differentrois} (Left), the results suggest that performing regional registration after applying the proposed algorithm does not always provide reasonable results for the small-sized ROIs for the experiments with $MSL = 10$. However, regional registration is robust against medium- and large-sized ROIs with $MLS=10$ and different ROI sizes with $MSL = 1$.
Fig.~\ref{fig:differentrois} (Right) compares the registration accuracy of the proposed registration method followed by fine registration with those of the whole tissue registration method (\cite{moles2014registration} 1 Round) and the regional version of this method (\cite{moles2014registration} 2 Rounds) using $MSL = 1$ and $MSL = 10$ for fine registration.  One can see that the registration outputs are more accurate and robust for larger ROI sizes and the algorithm performs better using $MSL = 1$.\\

\clearpage
%Fig. \ref{fig:vessels2} shows a 2D view of the other 10 blood vessels in our experiment that were selected for registration (first column). The resulting 3D reconstructed blood vessels after performing registration using the proposed algorithm are shown in the second and third column, respectively. The fourth and fifth column show 3D reconstruction of the registered lumen masks using the proposed algorithm followed by fine registration, and regional version (\cite{moles2014registration} 2 rounds) of the method by~\cite{moles2014registration} as the registration technique.

\clearpage

\bibliographystyle{plain}
\bibliography{Main}